\documentclass[aps,prl,twocolumn,showpacs]{revtex4}
\usepackage{bm}
\usepackage{graphicx}

\begin{document}

\title{Friction in inflaton equations of motion}

\author{Ian D Lawrie}
\email[]{i.d.lawrie@leeds.ac.uk}
\affiliation{Department of Physics and Astronomy, The University of Leeds,
Leeds LS2 9JT, England}

\date{\today}

\begin{abstract}
The possibility of a friction term in the equation of motion for a scalar field is investigated
in non-equilibrium field theory. The results obtained differ greatly from existing estimates
based on linear response theory, and suggest that dissipation is not well represented by a
term of the form $\eta\dot{\phi}$.
\end{abstract}


\maketitle

Most realizations of inflationary cosmology suppose that at some point in the
universe's early history, its energy content became dominated by a classical
scalar field (the inflaton $\phi(t)$), whose equation of motion has the form
\begin{equation}
\ddot{\phi}+3(\dot{a}/a)\dot{\phi}+\eta(\phi)\dot{\phi}+V'_{\mathrm{eff}}(\phi)=0,
\label{eom1}
\end{equation}
where $a(t)$ is the cosmic scale factor. The friction term
$\eta(\phi)\dot{\phi}$ is important, because it provides a mechanism whereby the
energy of the inflaton may be converted into the matter with which the present
universe is filled.  It has also been speculated \cite{berera95,berera00} that
sufficiently large frictional effects might exercise a controlling influence on
the inflationary process itself. To derive this term from fundamental physics
is, however, not straightforward.  In fact, we argue in this letter that existing
estimates of the coefficient $\eta(\phi)$ are incorrect, and that dissipation
probably cannot be represented by a term $\eta\dot{\phi}$ at all.

The most systematic derivations to be found in the literature rely on linear response
theory \cite{morikawa85} (see also \cite{morikawa84,hosoya84,gleiser94,berera01}).  To be
concrete, suppose that $\phi(t)$ is the expectation value of a quantum field $\Phi(t,\bm{x})$
with the self-interaction $(\lambda/4!)\Phi^4$ and write $\Phi(t,\bm{x})=\phi(t)+\chi(t,\bm{x})$.
It is sufficient (see below) to deal with the Minkowski-space version of this theory, for which the
exact equation of motion is
\begin{equation}
\ddot{\phi}(t) + {\textstyle\frac{1}{2}}\lambda\langle\chi^2(t,\bm{x})\rangle\phi(t)+\ldots=0.
\label{eom2}
\end{equation}
We hope to extract from the expectation value $\langle\chi^2(t,\bm{x})\rangle$ a contribution proportional
to $\dot{\phi}$, and the ellipsis indicates terms that are irrelevant for this purpose. The particles
associated with the quantum field $\chi(t,\bm{x})$ have a time-dependent effective mass
$m^2(t)=m_\Phi^2 + \frac{1}{2}\lambda\phi^2(t)$ and, correspondingly, a time-dependent
energy $\omega_k(t)=\sqrt{k^2+m^2(t)}$, with $\dot{\omega}_k = \lambda\phi\dot{\phi}/2\omega_k$.
To use linear response theory, we assume that $\phi(t)$ varies slowly on a time scale set by
the relaxation times $\Gamma_k^{-1}$ that characterize the evolution of the state of the $\chi$
particles.  If we wish to evaluate the equation of motion at time $t$, when $\phi$ has the value
$\phi_0$, say, then at times $t'$ not too different from $t$ we write $\phi(t') = \phi_0 + \delta\phi(t')$
and treat $\delta\phi(t')\approx\dot{\phi}(t)(t'-t)$ as a small correction.  The Hamiltonian for $\chi$
is $H(t') \approx H_0+\delta H(t')$, where $H_0$ is the constant Hamiltonian obtained by setting
$\phi(t) = \phi_0$ and $\delta H(t') = (\lambda/2)\delta\phi(t')\int d^3x\left[\chi^2+\frac{1}{3}\chi^3\right]$.
We obtain an approximation to $\langle\chi^2(t,\bm{x})\rangle$ of
the Kubo type, namely
\begin{equation}
\langle\chi^2(t,\bm{x})\rangle\approx\langle\chi^2(t,\bm{x})\rangle^{\mathrm{eq}}
+i\int_{-\infty}^tdt'\langle\left[\delta H(t'),\chi^2(t,\bm{x})\right]\rangle^{\mathrm{eq}},
\label{kubo}
\end{equation}
where the expectation values are taken in the equilibrium state associated with the Hamiltonian $H_0$,
at a fixed temperature $\beta^{-1}$.  The second term, proportional to $\dot{\phi}$, can be calculated by
the methods of equilibrium perturbation theory \cite{lebellac00}.  To obtain a finite result,
one must approximately resum self-energy insertions, which have an imaginary part $\Gamma_k$.  Assuming that
$\Gamma_k\ll\omega_k$, the friction coefficient is
\begin{equation}
\eta(\phi_0) = \frac{\lambda^2\phi_0^2}{8}\int\frac{d^3k}{(2\pi)^3}\left[
\frac{\beta n_k^{\mathrm{eq}}(1+n_k^{\mathrm{eq}})}{\omega_k^2\Gamma_k}
+\frac{\Gamma_k(1+2n_k^{\mathrm{eq}})}{\omega_k^5}\right]
\label{etalr}
\end{equation}
where $n_k^{\mathrm{eq}}(t) = \left[\exp(\beta\omega_k(t))-1\right]^{-1}$ is the Bose-Einstein
distribution function.

Two heuristic arguments \cite{morikawa84,hosoya84} indicate the origins of the terms in (\ref{etalr}).
Morikawa and Sasaki \cite{morikawa84} treat $\chi(t,\bm{x})$ in the first instance as a {\it free} field.
It can be expanded in terms of annihilation and creation operators $a_k(t)$ and $a_k^\dag(t)$, chosen to
diagonalize the free-field Hamiltonian at time $t$, whose time dependence arises, in the usual Bogoliubov formalism,
from the time-dependent mass $m(t)$ -- an effect conventionally described as `particle creation'.  We
define two functions $n_k(t)$ and $\nu_k(t)$ through
\begin{eqnarray}
\langle a_k^\dag(t)a_{k'}(t)\rangle&=&(2\pi)^3\delta(\bm{k}-\bm{k}')n_k(t)\label{nkdef}\\
\langle a_k(t)a_{k'}(t)\rangle&=&(2\pi)^3\delta(\bm{k}+\bm{k}')\nu_k(t).\label{nukdef}
\end{eqnarray}
Of these, $n_k(t)$ (equal to $n_k^{\mathrm{eq}}(t)$ in a state of local equilibrium)
counts the number of particles per unit volume with momentum $k$, these `particles' being defined as the
quanta annihilated by $a_k(t)$, while $\nu_k(t)$ measures the extent to which the density
matrix at time $t$ is off-diagonal in the Fock basis associated with this definition of a particle.
The expectation value of interest is given by
\begin{equation}
\langle\chi^2(t,\bm{x}))\rangle
=\int\frac{d^3k}{(2\pi)^32\omega_k}\left[1+2n_k+2\mathrm{Re}\,\nu_k\right],
\label{chisqev}
\end{equation}
while $n_k(t)$ and $\nu_k(t)$ obey the exact evolution equations
\begin{eqnarray}
\partial_tn_k(t)&=&\frac{\dot{\omega}_k(t)}{\omega_k(t)}\mathrm{Re}\,\nu_k(t)\label{dndtI}\\
\partial_t\nu_k(t)&=&-2i\omega_k(t)\nu_k(t)+\frac{\dot{\omega}_k(t)}{2\omega_k(t)}
\left[1+2n_k(t)\right].\label{dnudtI}
\end{eqnarray}
It is argued in Ref.~\cite{morikawa84} that interactions introduce an imaginary part to the self-energy,
in effect replacing $\omega_k$ with $\omega_k-i\Gamma_k$.  Then, with the further assumption that all the
functions in (\ref{dnudtI}) are approximately constant over a time interval of order $\Gamma_k^{-1}$, this
equation can be integrated, yielding a contribution to $\nu_k(t)$ that reproduces the second term of
(\ref{etalr}), except that $n_k(t)$ need not have its equilibrium value.

Hosoya and Sakagami \cite{hosoya84} identify the first term of (\ref{etalr}) as arising from $n_k(t)$
in (\ref{chisqev}) by assuming that the evolution of this function is described by the kinetic equation
\begin{equation}
\partial_tn_k(t) = -2\Gamma_k\left[n_k(t)-n_k^{\mathrm{eq}}(t)\right],
\label{rtb}
\end{equation}
sometimes called the relaxation-time approximation to the Boltzmann equation.  On writing
$n_k(t) = n_k^{\mathrm{eq}}(t) + \delta n_k(t)$, with $\partial_tn_k^{\mathrm{eq}} = -\beta\dot{\omega}_k
n_k^{\mathrm{eq}}(1+n_k^{\mathrm{eq}})$, and assuming again that everything changes sufficiently slowly
on a time scale of $\Gamma_k^{-1}$, we obtain a solution for $\delta n_k$ which, when inserted into
(\ref{chisqev}), reproduces the first term of (\ref{etalr}).

The {\it ad hoc} features of these heuristic arguments (specifically, the fact that equations (\ref{dndtI})
and (\ref{dnudtI}) apply only to a free field theory, while the kinetic equation (\ref{rtb}) is essentially
a guess) are largely avoided by the linear response calculations based on (\ref{kubo}), developed by the
same authors.  However, the evaluation of the expectation values in (\ref{kubo}) relies in an essential
way on analytic properties of thermal Green functions (arising from the so-called KMS condition) which apply
only to a state of {\it exact} thermal equilibrium, and have no analogue for a system that departs, even to
a small extent, from equilibrium.  Since the equation of motion (\ref{eom1}) implies a non-equilibrium state
in which $\phi$ changes with time, it is important
to know whether the result (\ref{etalr}) applies to the slow-evolution limit of a non-equilibrium state.  If
it does, it might also serve as a useful approximation when the evolution is not especially slow.  (Even in
equilibrium, the expression (\ref{etalr}) may be unreliable owing to large corrections from higher orders of
perturbation theory \cite{jeon93,jeon95,jeon96}, but we do not address that issue here.)

To investigate this question, we have derived improved versions of equations (\ref{dndtI}) - (\ref{rtb})
which, though approximate, take systematic account both of interactions and of non-equilibrium time
evolution.  They can be expressed as
\begin{eqnarray}
\epsilon\partial_tn_k&=&\alpha_k-\Gamma_k(1+2n_k)
+\epsilon\frac{\dot{\omega}_k}{\omega_k}\mathrm{Re}\,\nu_k\label{dndt}\\
\epsilon\partial_t\nu_k&=&-2i(\omega_k-i\Gamma_k)\nu_k-\alpha_k+\epsilon\frac{\dot{\omega}_k}{2\omega_k}
(1+2n_k),\qquad
\label{dnudt}
\end{eqnarray}
where $\alpha_k$ is a function whose meaning will be explained shortly, while $\epsilon$, which has the value
1, has been introduced as a formal parameter to generate a time-derivative expansion.

Details of the somewhat lengthy derivation of these equations will be reported elsewhere, but the essential
strategy is the following.  We deal with Green functions
$\rho(t,\bm{x};t',\bm{x}') = i\langle[\chi(t,\bm{x}),\chi(t',\bm{x}')]\rangle$ and
$C(t,\bm{x};t',\bm{x}') = \frac{1}{2}\langle\{\chi(t,\bm{x}),\chi(t',\bm{x}')\}\rangle$, where $[\ldots]$ denotes the
commutator and $\{\ldots\}$ the anticommutator.  In particular, the expectation value that we wish to evaluate is
given by
\begin{equation}
\langle\chi^2(t,\bm{x})\rangle = C(t,\bm{x};t,\bm{x}).
\end{equation}
These Green functions obey exact Dyson-Schwinger equations, which are
conveniently expressed (see, e.g., Ref. \cite{aarts01}) in terms of self-energies $\Sigma_\rho$ and $\Sigma_C$ as
\begin{eqnarray}
[\partial_t^2+\omega_k^2(t)]\rho(t,t';\bm{k})&=&-\int_{t'}^t\!dt''\,\Sigma_\rho(t,t'';\bm{k})\rho(t'',t';\bm{k})
\nonumber\\
&&\label{rhoeqn}\\
{[}\partial_t^2+\omega_k^2(t)]C(t,t';\bm{k})&=&-\int_0^t\!dt''\,\Sigma_\rho(t,t'';\bm{k})C(t'',t';\bm{k})
\nonumber\\
&&+\int_0^{t'}\!dt''\,\Sigma_C(t,t'';\bm{k})\rho(t'',t';\bm{k})\nonumber\\\label{ceqn}
\end{eqnarray}
after a spatial Fourier transform.  Our basic approximation is to introduce local {\it ans\"atze} for
the self-energies.  First, the commutator function $\rho(t,t';\bm{k})$ reduces in thermal equilibrium to the
spectral density, whose temporal Fourier transform can reasonably (though not exactly) be represented by a
Breit-Wigner function, characterized by a thermal width $\Gamma_k$.  The corresponding non-equilibrium approximation
is equivalent to the assumption
\begin{equation}
\Sigma_\rho(t,t';\bm{k})\approx 2\left[\Gamma_k(t)\Gamma_k(t')\right]^{1/2}\partial_t\delta(t-t').
\label{sigmaans}
\end{equation}
Then, given that correlations decay roughly exponentially with time, on the time scale set by $\Gamma_k(t)$, it
is reasonable to approximate $\Sigma_C$, which is symmetric in its time arguments, as
\begin{equation}
\Sigma_C(t,t';\bm{k})\approx -2\omega_k(t)\alpha_k(t)\delta(t-t').\label{sigmacans}
\end{equation}
The functions $\Gamma_k(t)$ and $\alpha_k(t)$ remain to be determined.  On substituting these
{\it ans\"atze} into the Dyson-Schwinger equations, we obtain a pair of local differential equations for
approximate Green functions, say $\hat{\rho}$ and $\hat{C}$.  These equations can be solved in terms of
auxiliary functions $n_k(t)$ and $\nu_k(t)$ that satisfy the evolution equations (\ref{dndt}) and (\ref{dnudt}).
In particular, $\hat{C}(t,\bm{x};t,\bm{x})$ is then given precisely by the expression (\ref{chisqev}).  In the
interacting theory, though, one cannot identify $n_k(t)$ and $\nu_k(t)$ unambiguously in terms of the
expectation values (\ref{nkdef}) and (\ref{nukdef}).

The approximate Green functions obtained in this way can be taken to serve as the lowest-order propagators
in a partially-resummed perturbation theory, similar to that described in Ref. \cite{lawrie89}. The result
of substituting the solutions of (\ref{dndt}) and (\ref{dnudt}) into the expression (\ref{chisqev}) is the
expectation value $\langle\chi^2(t,\bm{x})\rangle$ evaluated at the lowest order of this perturbation
theory.  Here, we shall be satisfied with this lowest-order estimate, though the approximation can in
principle be pursued to arbitrarily high orders. Equally, this perturbation theory can be used to evaluate
the self-energies in (\ref{sigmaans}) and (\ref{sigmacans}).  With a suitable prescription for extracting
local approximations to these self-energies, we obtain concrete expressions for the functions $\alpha_k(t)$
and $\Gamma_k(t)$.  At 2-loop order, we find
\begin{eqnarray}
\alpha_k(t)&=&\frac{c}{\pi^2}\int d^3k_1d^3k_2d^3k_3
\frac{\Delta}{\omega_k\omega_{k_1}\omega_{k_2}\omega_{k_3}}
\nonumber\\
&&\times\left[(1+n_{k_1})(1+n_{k_2})n_{k_3}+n_{k_1}n_{k_2}(1+n_{k_3})\right]\nonumber\\
\label{alphares}\\
\Gamma_k(t)&=&\frac{c}{\pi^2}\int d^3k_1d^3k_2d^3k_3
\frac{\Delta}{\omega_k\omega_{k_1}\omega_{k_2}\omega_{k_3}}
\nonumber\\
&&\times\left[(1+n_{k_1})(1+n_{k_2})n_{k_3}-n_{k_1}n_{k_2}(1+n_{k_3})\right]\nonumber\\
\label{gammares}
\end{eqnarray}
where $c=\lambda^2\pi^2/64(2\pi)^5$ and $\Delta$ is the product of delta functions that conserve energy and
momentum in the 2-body collisions $(\bm{k}_1,\bm{k}_2)\rightleftharpoons(\bm{k}_3,\bm{k})$.

With these expressions for $\alpha_k$ and $\Gamma_k$, the evolution equation (\ref{dndt}) for $n_k(t)$ is just
what we might expect: a genuine Boltzmann equation in place of the approximate version hypothesized in
(\ref{rtb}).  The quantity $S_k=\alpha_k-\Gamma_k(1+2n_k)$ is the scattering integral that accounts for the net
rate of change of the number of particles with momentum $k$ due to collisions, while the remaining term is the
source term from (\ref{dndtI}) associated with particle creation.  An heuristic expectation for the
evolution of $\nu_k(t)$ is not easy to formulate.  We can observe in (\ref{dnudt}), though, that $\omega_k$ has
indeed been replaced with $\omega_k-i\Gamma_k$ in the first term of (\ref{dnudtI}), though not elsewhere.

We investigate the solution of (\ref{dndt}) and (\ref{dnudt}) in the limit of slow time evolution by expanding
in powers of the formal parameter $\epsilon$, so as to generate a time-derivative expansion.  For $n_k(t)$, the
leading term is just the Bose-Einstein distribution $n_k^{\mathrm{eq}}(t)$, which makes $S_k$ vanish.  The inverse
temperature $\beta$ is, of course, a free parameter specifying a particular state.  For $\nu_k(t)$, the leading
term  is
\begin{equation}
\nu_k^{\mathrm{eq}}(t)=\frac{i\alpha_k^{\mathrm{eq}}(t)}{2[\omega_k(t)-i\Gamma_k^{\mathrm{eq}}(t)]}
=\frac{i\Gamma_k^{\mathrm{eq}}(t)[1+2n_k^{\mathrm{eq}}(t)]}{2[\omega_k(t)-i\Gamma_k^{\mathrm{eq}}(t)]}
\end{equation}
where $\alpha_k^{\mathrm{eq}}(t)$ and $\Gamma_k^{\mathrm{eq}}(t)$ are the above expressions with
$n_k = n_k^{\mathrm{eq}}(t)$. The next-to-leading terms $\delta n_k(t)$ and $\delta\nu_k(t)$ are proportional
to $\dot{\phi}$, and it is from these that we can obtain $\eta(\phi)$ by substitution in (\ref{chisqev}). Here, we
use the auxiliary approximation that $\Gamma_k^{\mathrm{eq}}\ll\omega_k$. This slightly
simplifies the numerical calculations to which we must shortly resort, but we have adopted it primarily for
consistency with the strategy that leads to the linear-response result (\ref{etalr}), with which we plan to compare
our own. With this approximation, we find that
\begin{equation}
{\mathrm{Re}}\,\delta\nu_k = \frac{1}{2\omega_k^2}\left[\frac{\dot{\omega}_k}{2\omega_k}\Gamma_k^{\mathrm{eq}}
(1+2n_k^{\mathrm{eq}})-\omega_k\partial_t{\mathrm{Im}}\,\nu_k^{\mathrm{eq}}\right]
\label{deltanu}
\end{equation}
while $\delta n_k$ is the solution of the integral equation
\begin{equation}
\int_0^\infty dk'K(k,k')\delta n_{k'} = \partial_tn_k^{\mathrm{eq}}-\frac{\dot{\omega}_k}{\omega_k}
{\mathrm{Re}}\,\nu_k^{\mathrm{eq}},
\label{deltan}
\end{equation}
with $K(k,k') = \left.\delta S_k/\delta n_{k'}\right\vert_{n_k=n_k^{\mathrm{eq}}}$.

At this point, we observe that the linear-response result can be recovered by means of two
further approximations. The first is to set $K(k,k') = -2\Gamma_k^{\mathrm{eq}}\delta(k-k')$, which means
replacing $\alpha_k$ and $\Gamma_k$ in (\ref{dndt}) with their equilibrium values. The second is to take the
time dependence of $\Gamma_k(t)$ to be of the form $\Gamma_k(t)\approx{\mathrm{const.}}/\omega_k(t)$.  This
amounts to replacing the {\it ansatz} (\ref{sigmaans}) with
$\Gamma_k(t)\approx i\Sigma_\rho^{\mathrm{eq}}(\omega_k,\bm{k})/2\omega_k(t)$, and taking the equilibrium
self-energy to be constant.

\begin{figure}
\includegraphics[scale=0.5]{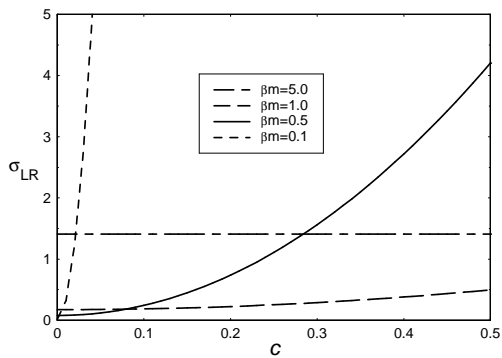}
\caption{\label{fig1}The friction coefficient $\sigma$ as calculated from the linear response formula
(\ref{etalr}).}
\end{figure}

\begin{figure}
\includegraphics[scale=0.5]{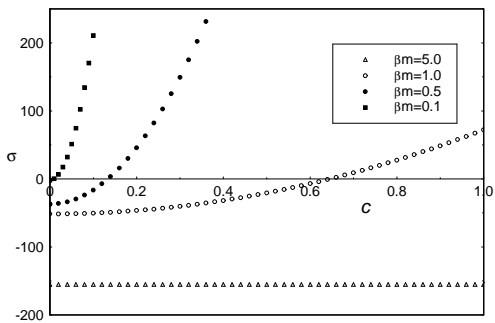}
\caption{\label{fig2}The friction coefficient $\sigma$ as calculated numerically from (\ref{deltanu}) and
(\ref{deltan}).}
\end{figure}

To avoid these extra approximations, we must, in particular, solve the integral equation (\ref{deltan}) using
the full kernel $K(k,k')$.  This equation is in fact not self-consistent. The reason is that the scattering
integral $S_k$ conserves both energy and (at 2-loop order where only elastic collisions are included) particle
number.  This implies two sum rules
\begin{equation}
\int_0^\infty dk k^2K(k,k') =\int_0^\infty dk k^2\omega_kK(k,k') =0,
\label{sumrules}
\end{equation}
which are not respected by the right-hand side of (\ref{deltan}).  In principle, the equation has no solution.
This means that the time-derivative expansion does not exist, and that dissipation cannot properly be
represented by a term proportional to $\dot{\phi}$ in the equation of motion. Nevertheless, we have obtained
numerical solutions to (\ref{deltan}) that are rather well defined.  This is possible because the right-hand
side becomes very small as $k$ increases beyond quite modest values and the same turns out to be true of the
solution $\delta n_k$.  In effect, the solution involves only a truncated kernel, which is not constrained
by the sum rules.  The existence of this approximate solution allows us to construct an effective friction
coefficient, and thus to quantify the effect of the extra approximations that lead to (\ref{etalr}). The effect
is rather large, as indicated in figures \ref{fig1} and \ref{fig2}. Figure \ref{fig1} shows the dimensionless
quantity $\sigma=\eta(\phi)m/(128\pi\phi^2)$, calculated from the linear response formula (\ref{etalr}) as a
function of the coupling strength $c$ defined above, for selected values of the dimensionless parameter $\beta m$.
The haphazard appearance of the four curves, which have essentially the same shape when viewed on appropriate
scales, is explained by a competition between the two terms in (\ref{etalr}). Since $\Gamma_k$ is
proportional to $c$, weak coupling favours the first term, which is a decreasing function of temperature,
while strong coupling favours the second term, which increases with temperature.
Figure \ref{fig2} shows $\sigma$ as calculated numerically from equations (\ref{deltanu}) and (\ref{deltan}).
Not only does the magnitude differ by factors of several hundred, but $\sigma$ turns out to be negative at weak
coupling.  (Strictly, it is only at weak coupling that the assumption $\Gamma_k\ll\omega_k$ holds.) Of course, a
negative friction coefficient does not make good physical sense:  it clearly reflects the breakdown of the
time-derivative expansion noted above.

We estimate the accuracy of the numerical integrals used to obtain $K(k,k')$ and to evaluate (\ref{chisqev}) as
about 1\%, and have verified the sum rules (\ref{sumrules}) at this level.  The solution of (\ref{deltan}) for
$\delta n_k$ is such that the left and right-hand sides of this equation differ by no more than one part in
$10^7$ (i.e. the solution is valid to full single-precision accuracy) and we have verified that the
whole procedure converges to the results shown as the momentum cutoff used in (\ref{chisqev}) and
(\ref{deltan}) is increased.

These results strongly indicate that linear response theory gives the wrong answer to this problem, and
the reason may be this.  The formula (\ref{etalr}) arises from the second term in (\ref{kubo}) where,
for the mode of momentum $k$, the time integral amounts to an average over a time interval of order $\Gamma_k^{-1}$.
During this time interval, the actual state of the system changes by an
amount of order $\Gamma_k^{-1}\dot{\phi}$, so the putative equilibrium state represented by
$\langle\ldots\rangle^{\mathrm{eq}}$ is uncertain by an amount of this order.  Thus, the {\it first} term of
(\ref{kubo}) is ill-defined by an amount proportional to $\dot{\phi}$ that could well be comparable with, or
greater than, the quantity we extract from the second.

Two final remarks are in order. First, the field theory in an expanding universe is equivalent, in conformal time,
to a Minkowski-space theory with a mass $m_\Phi$ that depends on $a(t)$.  Applied to this theory, the time-derivative
expansion generates additional terms, proportional to $\dot{a}/a$, which may be important, but do not affect our
conclusions about the term proportional to $\dot{\phi}$.  Second, the results presented here may be specific to the
$\lambda\Phi^4$ model.  The dynamics of fermions to which $\Phi$ might couple can be represented \cite{lawrie00}
by evolution equations similar to (\ref{dndt}) and (\ref{dnudt}), but $K(k,k')$ and $\Gamma_k$ are different, and
the solution may be very sensitive to this difference.

It is a pleasure to acknowledge discussions with Arjun Berera and Rudnei Ramos, which prompted the study
reported here, and to thank Jim Morgan for valuable comments on an earlier version of this paper.


\end{document}